\documentclass{SVProc}
\usepackage{url}

\usepackage{graphicx}
\usepackage{amsmath}
\usepackage{amssymb}
\usepackage{fancyhdr}
\usepackage[colorlinks,urlcolor=blue,linkcolor=blue,citecolor=blue]{hyperref}
\fancypagestyle{preprint}{\fancyhf{} \fancyfoot[L]{*\textit{Preprint version of the paper submitted to the 8th World Conference on Structural Control and Monitoring (8WCSCM)}}}

\fancypagestyle{github}{\fancyhf{} \fancyfoot[L]{*\url{https://github.com/mahindrautela/anoVAE-DelamSHM}}}

\begin{document}
\mainmatter              
\title{Deep generative models for unsupervised delamination detection using guided waves}
\titlerunning{Deep generative models for unsupervised delamination detection...}  
%
\author{Mahindra Rautela\inst{1,2} \and Amin Maghareh\inst{2}
Shirley Dyke\inst{2} \and S. Gopalakrishnan\inst{1}}
\authorrunning{M. Rautela et al.} 
%
%
\institute{Department of Aerospace Engineering, Indian Institute of Science, Bangalore\\
\email{mrautela@iisc.ac.in}
\and
School of Mechanical Engineering, Purdue University, IN 47907-2088, USA}

\maketitle              

\begin{abstract}
With the rising demands for robust structural health monitoring procedures for aerospace structures, the scope of intelligent algorithms and learning techniques is expanding. Supervised algorithms have shown promising results in the field provided a large, balanced, and labeled amount of data for training. For some applications like aerospace, the data collection process is cumbersome, time-taking, and costly. Also, generating possible damage scenarios in a laboratory setup is challenging because of the complexity of the damage initiation and failure mechanism. Besides this, the uncertainties of the real-time operation restrict the online prediction accuracy with supervised learning. In this paper, deep generative models are proposed for unsupervised delamination prediction as an anomaly detection problem. In this one-class-based model, the deep learning network is trained to learn the distribution of baseline signals. In the testing phase, damage signals and unseen baseline signals are fed into the trained network to predict the state of the structure, i.e., healthy or unhealthy (delamination). It is seen that the proposed method can successfully predict the delamination with high accuracy.

\end{abstract}
\thispagestyle{preprint}

\section{Introduction}
Composite materials fail via complex mechanisms, and delaminations are the most common ones which need proper attention. Condition-based monitoring can eliminate costly and time-taking scheduled maintenance procedures \cite{rautela2021inverse}. Ultrasonic guided wave is one of the popular techniques to actively interrogate the composite structures to identify the delaminations \cite{mitra2016guided}. Recently, data-driven supervised deep learning algorithms have shown promising results for damage identification using ultrasonic guided waves \cite{rautela2021combined}. However, the well-labeled data-collection process is cumbersome, costly, and time-taking. It is difficult to obtain data for possible damage scenarios in some applications. Due to this, the datasets become unbalanced, which restricts the use of supervised deep learning algorithms. In such scenarios, damage detection can be posed as an unsupervised one-class anomaly detection problem. In this framework, the network is trained to learn the distribution of the pristine state of the structure. In testing phase, the out-of-distribution signals are classified as damages \cite{rautela2022delamination}.

Deep generative models are one of the advance unsupervised learning techniques aimed at capturing the probabilistic governing distribution that generates a class of data to generate similar data \cite{oussidi2018deep}. Deep  generative models can be divided into two main categories: 
\begin{enumerate}
	\item Cost-function based models like Variational Autoencoders (VAE) \cite{kingma2013auto} and Generative Adversarial Networks (GAN) \cite{goodfellow2014generative}.
	\item Energy-based models like Deep Boltzmann Machines \cite{hinton2012better} and Deep Belief Networks \cite{hinton2006fast}.
\end{enumerate}

Out of different deep generative models and their variants, VAE and GAN are most popularly used for anomaly detection. In the context of one-class infrastructure health monitoring problems, limited research investigations are carried out in the literature. VAE is used for damage detection in bridges \cite{ma2020structural}, rail squats localization \cite{yuan2021unsupervised} and damage detection in tunnels \cite{zhang2022unsupervised}. GAN is used for road damage detection \cite{maeda2021generative}, post-disaster building damage detection \cite{tilon2020post}, bridges health monitoring \cite{jiang2021continuous}, concrete cracking and spalling \cite{gao2021balanced}. Most of the literature is inclined toward civil-related infrastructures with little to no investigations of aerospace structures. Besides this, vibration-based or vision-based sensing technologies are prominently used in the literature. 

In this work, we have focused on the one-class version of VAE for delamination detection in aerospace composite panels. The ultrasonic guided waves are used to interrogate the panel using a sensor array. The raw time-series guided wave signals are converted into time-frequency-based higher-order representations for training purposes. The paper is presented as follows: Section-\ref{sec:theory} contains theoretical background on variational autoencoders, Section-\ref{sec:mm} consists of the description of the dataset and training procedures, Section-\ref{sec:results} presented the results of delamination detection using VAE and the paper is concluded in Section-\ref{sec:conclusion}.

\section{Theoretical Background} \label{sec:theory}
\subsection{Variational Autoencoders} \label{ssec:vaetheory}
\begin{figure}[h!]
	\centering
	\begin{minipage}[b]{0.48\textwidth}
		\centering
		\includegraphics[width=0.85\textwidth]{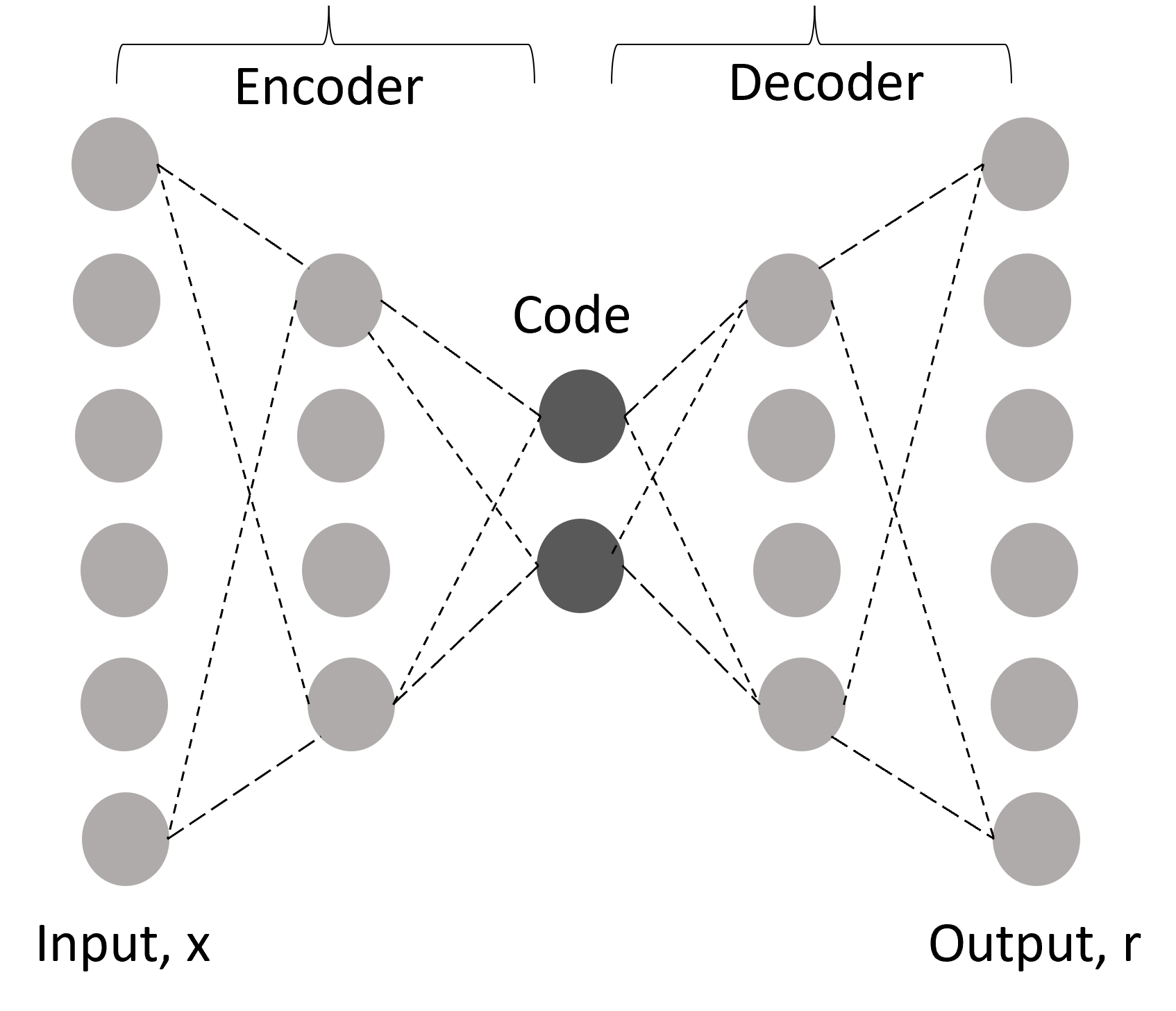}
	\end{minipage}
	\begin{minipage}[b]{0.48\textwidth}
		\centering
		\includegraphics[width=1.0\textwidth]{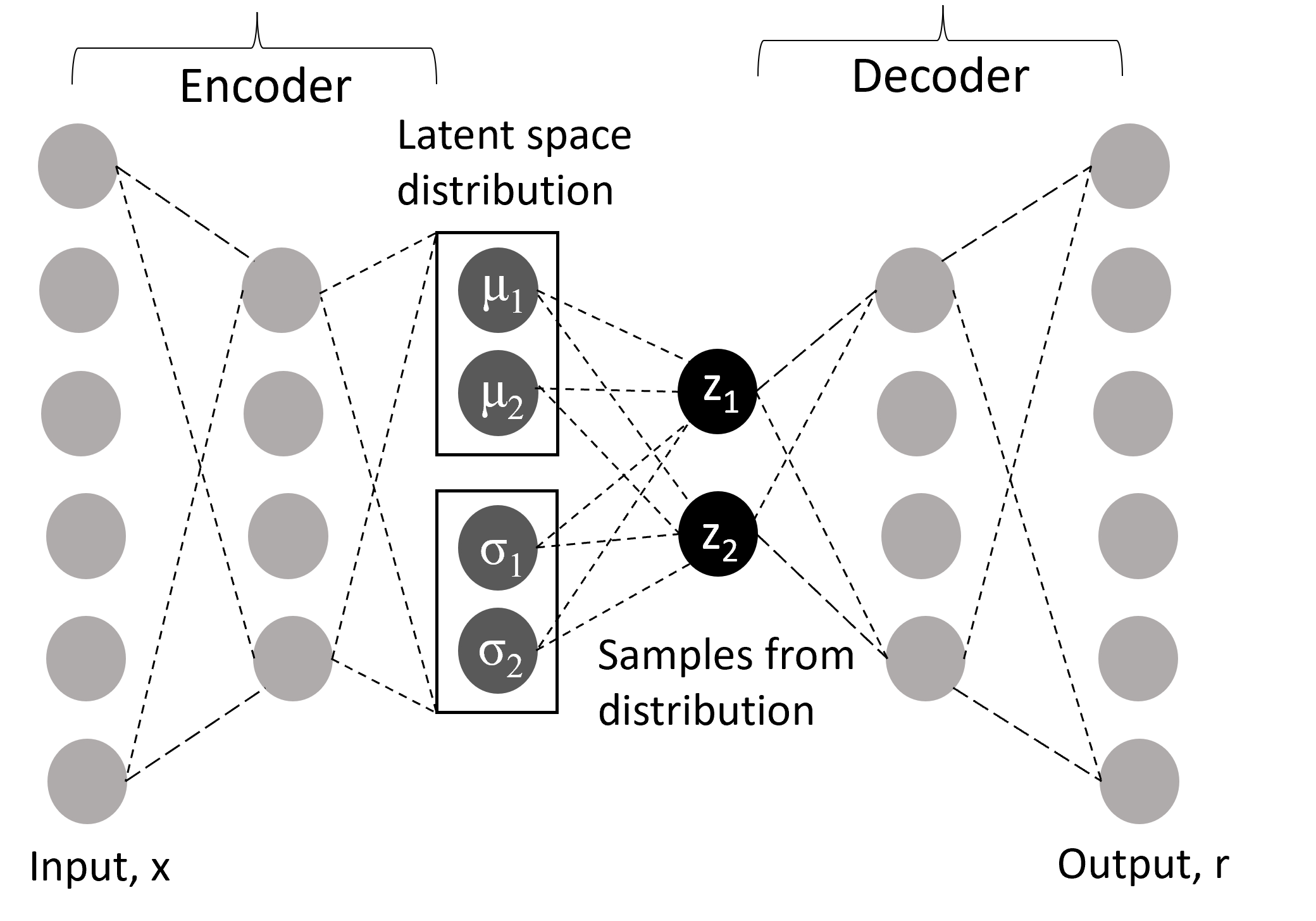}
	\end{minipage}
	\caption{General architecture of (a) Basic autoencoder (b) Variational autoencoder}
	\label{fig:ae}
\end{figure}

A basic autoencoder has three components, i.e., encoder, latent space (or code/hidden representation), and decoder, as shown in Figure-\ref{fig:ae}(a). An encoder compresses the input into latent space, and a decoder reconstructs backs the input from the latent space. During this process, the latent layer learns valuable features of the dataset. 

Variational autoencoder (VAE) is similar to basic autoencoder, but it provides a probability distribution instead of a single value for each latent variable. The decoder is fed with a randomly generated vector from the latent state distribution. In addition to regularization, this helps VAE to enforce a continuous and smooth latent space representation \cite{kingma2013auto}. Mathematically, $p(x,z)$ needs to be calculated to infer the latent space using Bayesian inference, which is represented by Eq.\ref{eq:bayes}. The term $\int p(x|z) p(z) dz$ represents an intractable distribution, so Eq.\ref{eq:bayes} cannot be solved directly \cite{zhang2019semi}.

\begin{equation} \label{eq:bayes}
	p(z|x) = \frac{p(x|z) p(z)}{\int p(x|z) p(z) dz}
\end{equation}

However, variational inference is used to estimate the integral. For this, the $p(z|x)$ distribution is approximated with $q(z|x)$ in such a way that it can be a tractable distribution. To ensure both distributions are similar, Kullback-Leibler (KL) divergence is minimized, $min \hspace{1mm} D_{KL}(q(z|x)||p(z|x))$. This expression can be minimized by maximizing $E_{q(z|x)} \log p(x|z) - D_{KL}(q(z|x)||p(z))$ \cite{pol2019anomaly}. The first term is the reconstruction likelihood which is similar to reconstruction error between input and output and the second term imposes similarity between prior and posterior distribution for latent variables. The prior distribution is assumed to be standard Normal distribution $N(0,I)$ and posterior is estimated of the form $N(\mu_{\phi}(x),\sigma^2_{\phi}(x))$. To implement it in backpropagation based  neural-network framework, reparameterization trick is performed \cite{kingma2019introduction}. The overall loss function is the summation of reconstruction error and the KL divergence.

\subsection{One-class Variational Autoencoder}
\begin{figure}[h!]
	\centering
	\includegraphics[width=0.85\textwidth]{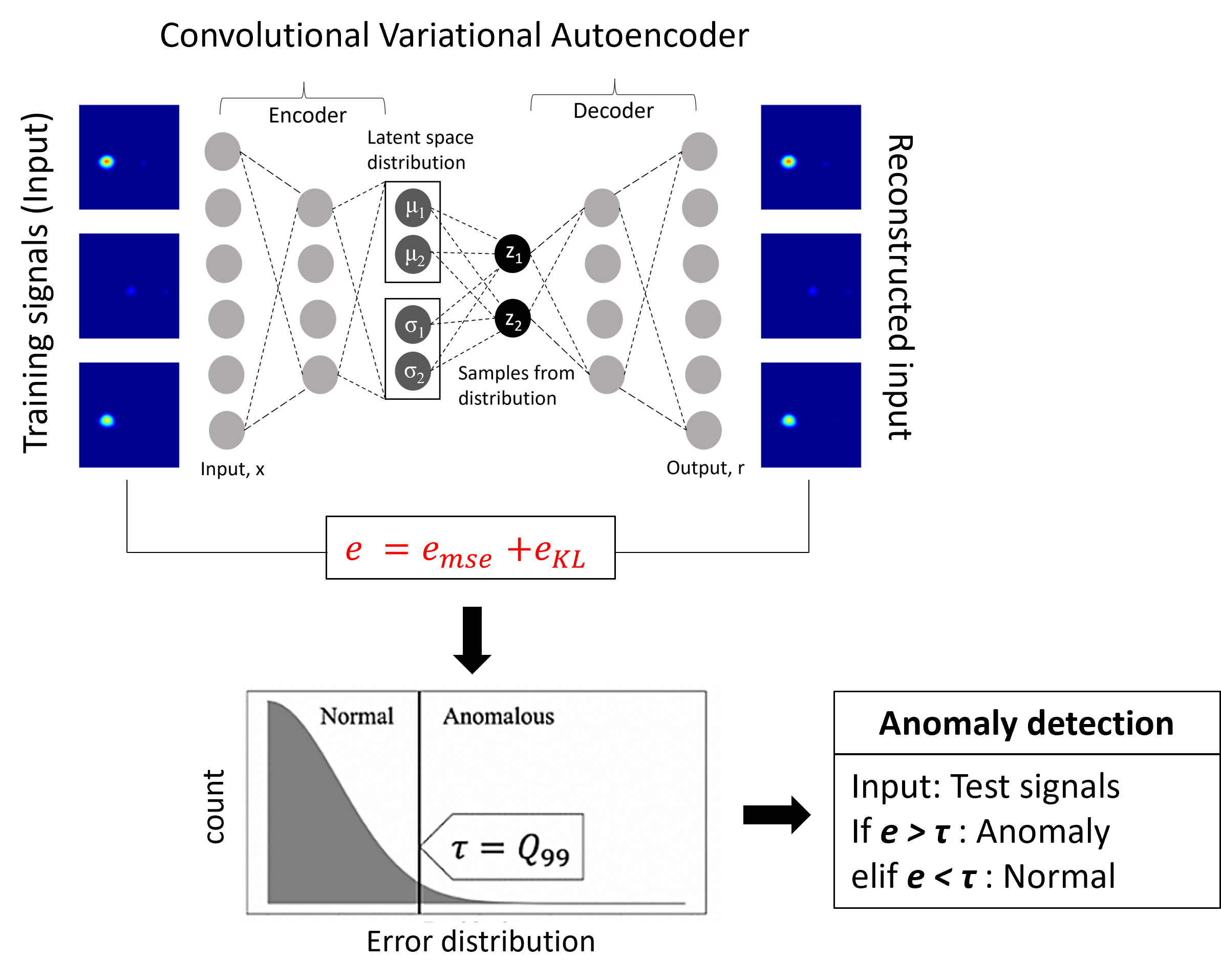}
	\caption{Delamination detection using VAE and thresholds. If error is greater than the threshold, the signal is classified as anomaly (delamination).}
	\label{fig:strategyVAE}
\end{figure}

In this work, a one-class version of VAE is used for delamination detection. Here, the network is trained to learn the distribution of baseline signals. In the testing phase, unseen baseline and delamination signals are used to test the network. The proposed methodology belongs to a broader class of anomaly detection problems. The strategy for unsupervised delamination detection using VAE is shown in Fig.~\ref{fig:strategyVAE}. In this strategy, the errors (mean squared reconstruction + KL divergence) are collected after the training process. Based on the distribution of errors, thresholds are selected to detect the delaminations. In this work, the 99th quartile and maximum value of errors are used as two different thresholds. Selection of an optimum threshold depends on uncertainties of real-time operation, and application-dependent safety-cost tradeoffs \cite{rautela2022delamination}. This investigation is not in the scope of the current work.

\section{Materials \& Methods} \label{sec:mm}
\subsection{Dataset}
\begin{figure}[h!]
	\centering
	\begin{minipage}[b]{0.48\textwidth}
		\centering
		\includegraphics[width=1.0\textwidth]{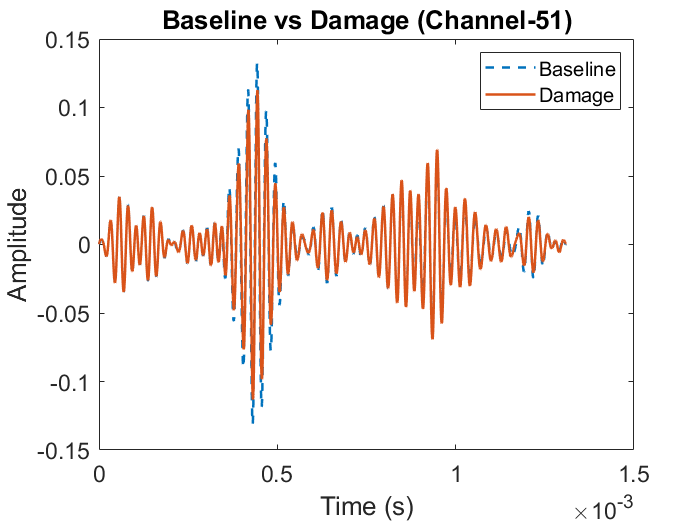}
	\end{minipage}
	\begin{minipage}[b]{0.48\textwidth}
		\centering
		\includegraphics[width=1.0\textwidth]{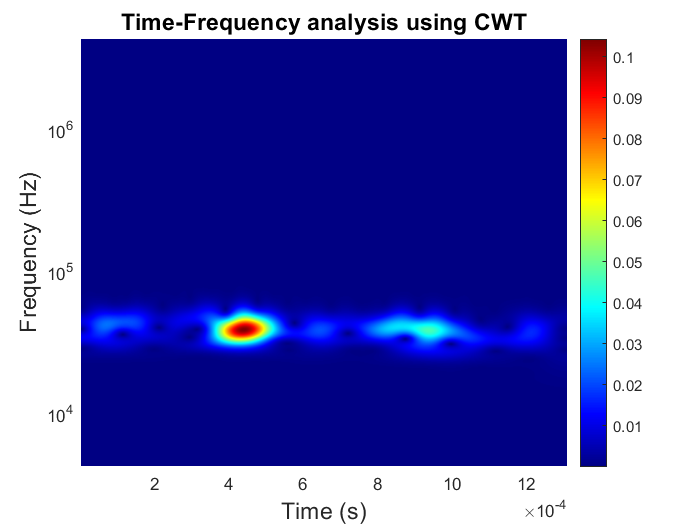}
	\end{minipage}
	\caption{Experimental dataset (a) Raw guided wave signals (Baseline and Damage), (b) CWT representation of the baseline signal.}
	\label{fig:ogwplots}
\end{figure}

In this work, an experimental guided-wave benchmark dataset from the Open Guided Waves platform is used \cite{moll2019open}. The experiments are performed on a 500 $\times$ 500 $\times$ 2 mm T700M21 carbon fiber reinforced polymer with a quasi-isotropic layup placed in a climatic chamber. Twelve piezoelectric transducers are arranged, giving rise to 66 signals per experiment. Twelve different frequencies ranging from 40 kHz to 260 kHz are used for excitation. Damages at 28 different positions are introduced using a reversible damage model. The wavefield analysis shows that the reversible damage model behaves like real delamination. The original dataset contains 47,520 baseline and 22,176 damage signals. 5000 baseline and damage signals are randomly picked for the training and testing purpose. This random selection also includes new samples picked every time the algorithm runs. In this work, the 1d raw guided-wave signals are converted into 2d time-frequency representations using continuous wavelet transformation (CWT) \cite{rautela2022delamination}. Mathematically, CWT of a function $F(t)$ is represented by $F^W$ \cite{rautela2021ultrasonic}, shown by Eq. \ref{eq:cwt}. 

\begin{equation}\label{eq:cwt}
	F^{W}(a,b) = \int_{-\infty}^{+\infty} F(t) \Phi \Bigg(\frac{t-b}{a}\Bigg)dt
\end{equation}

where, $\Phi$(t) is a wavelet basis function, 'a' and 'b' defines the width and position in time of $\Phi(t)$. Fig.~\ref{fig:ogwplots} shows time-series signals from the dataset and it's corresponding CWT representation.

\subsection{Training procedure}
The network is trained with the summation of mean squared reconstruction error and KL-divergence as the loss function (See Section-\ref{ssec:vaetheory}). The Adam optimization scheme with a learning rate of 1e-3 and a batch size of 32. The network consists of three parts, encoder, decoder, and latent space. Both encoder and decoder consist of convolutional layers with a kernel size of 3 × 3, filters: 16, 32, 64, 128, 256, a stride length of 2, and a Leaky ReLU activation function. The dimension of the latent space is 2. The code files are developed in Python programming language, and the code package is made open-source on GitHub*.

\thispagestyle{github}

\section{Results \& Discussions} \label{sec:results}
\begin{figure}[h!]
	\centering
	\includegraphics[width=0.95\textwidth]{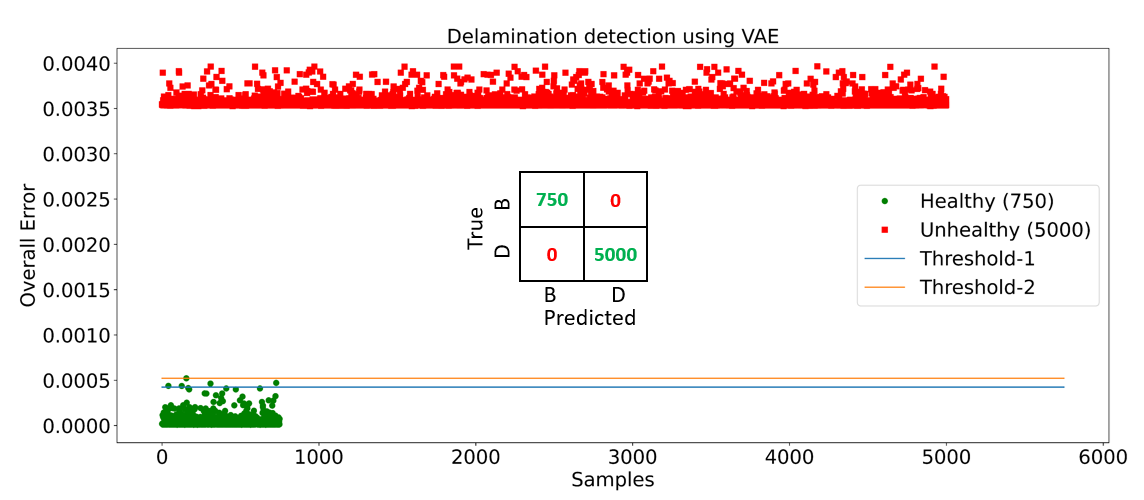}
	\caption{Delamination detection using VAE and thresholds. The confusion matrix shows the classification results of unseen healthy and unhealthy (delamination) signals.}
	\label{fig:vaethresholds}
\end{figure}

\begin{figure}[h!]
	\centering
	\includegraphics[width=0.5\textwidth]{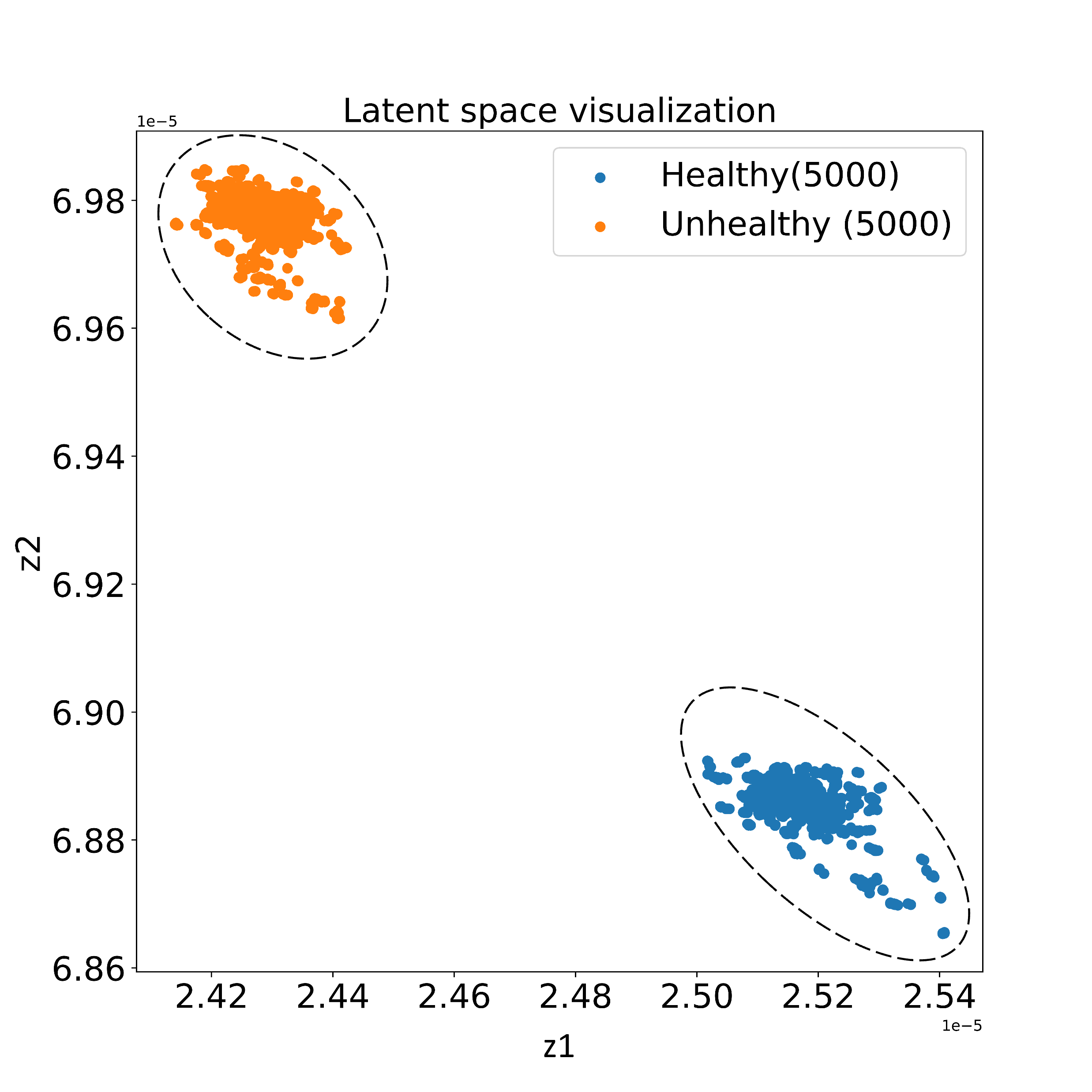}
	\caption{Two-dimensional latent space visualization of all healthy and unhealthy (delamination) signals.}
	\label{fig:vaelatent}
\end{figure}

After training the networks, the delamination detection strategy, as shown in Fig.~\ref{fig:strategyVAE} is implemented for testing. The errors are collected corresponding to the training samples. Two different thresholds are calculated, i.e., 99th quartile and maximum value of the error of the training samples. The trained network is fed with unseen healthy and delamination signals. If the error of the test sample exceeds the threshold, then it is classified as delamination. The results are plotted in Fig.~\ref{fig:vaethresholds}. It is seen that threshold-2 (maximum error) gives better classification accuracy. One of the reasons for such promising results is due to the use of higher-order wavelet enhanced representations instead of raw guided-wave signals \cite{rautela2022delamination}. The latent space distribution (mean value) for all the healthy and unhealthy signals is shown in Fig.~\ref{fig:vaelatent}. It can be seen that both baseline and delamination signals are well separated in the latent space. It is also observed that there exists a significant separation between the overall error of healthy and unhealthy signals (Fig.~\ref{fig:vaethresholds}). A similar distinction can be seen between healthy and unhealthy clusters in the latent space. (Fig.~\ref{fig:vaelatent}). For much-complicated problems and datasets, there may not be an evident separation between the distribution of baseline signals and anomalies \cite{rautela2022delamination}. In such cases, the selection of decision thresholds is decided based on the tradeoff between false positives rate (FPR) and false negatives rate (FNR). This tradeoff directly translates to the safety-cost tradeoff, which is application-specific. For safety-critical health monitoring applications like space habitats and nuclear power plants, reducing FPR is considered preferable to FNR.

\section{Conclusion} \label{sec:conclusion}
In this paper, a one-class version of the variational autoencoder is proposed for unsupervised delamination detection. The raw guided wave signals are converted to higher-order representations using continuous wavelet transformation. The network is trained to learn the distribution of baseline signals. Once the network is trained, thresholds are calculated and used to detect delaminations as out-of-distribution signals. This anomaly detection data-driven framework is helpful for data-scarce applications where balanced and labeled data collection is infeasible, costly, or time-taking. It is seen that the network provides promising results on the open-source benchmark dataset. This work will be extended and validated on different datasets in our upcoming work. Besides this, one-class GAN is another direction for future research investigations. The comparison between VAE and GAN capabilities to learn the baseline distribution will be an important highlight of our future work.

\bibliographystyle{splncs03.bst}
\bibliography{mybibfile}

\end{document}